\documentclass[12pt]{iopart}

\usepackage{amstext}
\usepackage{graphicx}
\usepackage{epstopdf}
\begin{document}


\title{Cavity-enhanced storage in an optical spin-wave memory}

\author{P. Jobez$^1$, I. Usmani$^1$\footnote{Present address: Laboratoire Charles Fabry, Institut d'Optique, CNRS, Universit\'{e} Paris-Sud, France}, N. Timoney$^1$, C. Laplane$^1$, N. Gisin$^1$, M. Afzelius$^1$}
\address{$^1$Group of Applied Physics, University of Geneva, Geneva, Switzerland}
\ead{mikael.afzelius@unige.ch}
\begin{abstract}
We report on the experimental demonstration of an optical spin-wave memory, based on the atomic frequency comb (AFC) scheme, where the storage efficiency is strongly enhanced by an optical cavity. The cavity is of low finesse, but operated in an impedance matching regime to achieve high absorption in our intrinsically low-absorbing Eu$^{3+}$:Y$_2$SiO$_5$ crystal. For storage of optical pulses as an optical excitation (AFC echoes), we reach efficiencies of 53\% and 28\% for 2$\mu$s and 10$\mu$s delays, respectively. For a complete AFC spin-wave memory we reach an efficiency of 12\%, including spin-wave dephasing, which is a 12-fold increase with respect to previous results in this material. This result is an important step towards the goal of making efficient and long-lived quantum memories based on spin waves, in the context of quantum repeaters and quantum networks.
\end{abstract}

\submitto{\NJP}
\maketitle

\section{Introduction}

Quantum information science seeks to develop methods and techniques that would allow us to perform certain tasks that cannot be achieved with conventional techniques based on classical physics. Entanglement-based quantum networks would play an important role in this perspective in order to provide quantum connectivity, distributed quantum computing \cite{Kimble2008} and long-distance quantum communication \cite{Sangouard2011}. An essential ingredient in quantum networks are quantum memories, which would locally store quantum states of light in the network. Quantum memories are explored using a wide range of systems and light-matter interaction schemes, see \cite{Hammerer2010,Simon2010,Bussieres2013} for recent reviews. A promising solid-state approach is based on rare-earth-ion doped crystals (REIC) \cite{Tittel2010b}, since these systems feature long optical and hyperfine coherence times at temperatures of 4K or below. The solid-state environment, however, causes inhomogeneous broadening of the optical lines, usually in the range of 100 MHz-10 GHz. A number of schemes have been proposed in order to control the inhomogeneous dephasing \cite{Moiseev2001,Kraus2006,Hetet2008,Afzelius2009a,Moiseev2011}, all based on photon-echo like processes, that in principle should allow low-noise operation at the single photon level (SPL). To date most reported work at the SPL using REIC memories have been based on the atomic frequency comb (AFC) method \cite{Riedmatten2008,Sabooni2010,Clausen2011,Saglamyurek2011,Gundogan2012,Zhou2012}. But these experiments have only implemented a partial version of that scheme, where the photon is stored solemnly as a optical excitation, called two-level AFC echo henceforth, without applying any optical-to-spin conversion, in which case the AFC memory is equivalent to a fixed optical delay line. The implementation of the optical-to-spin conversion is crucial for AFC quantum memories, called spin-wave AFC memory henceforth, because it allows to read out the memory on demand. Spin-wave storage could also result in unprecedented long storage lifetimes, at least in the milliseconds regime \cite{Alexander2007} and potentially in the range of minutes \cite{Heinze2013}, or longer.

Recently we demonstrated the first spin-wave AFC memory at the SPL \cite{Timoney2013}. The signal-to-noise ratio (SNR) was of the order of one for a few photons in the stored pulse. This is insufficient for storage of, for instance, quantum correlated photons. Our analysis \cite{Timoney2013} of the results clearly showed that the unconditional noise floor was sufficiently low for allowing operation at the SPL in principle, but the memory efficiency was insufficient due to the low optical depth of the REIC that we employed (a Eu$^{3+}$:Y$_2$SiO$_5$ crystal). The goal of this paper is to address the inefficiency of optical spin-wave memories based on weakly absorbing materials, in particular Eu$^{3+}$:Y$_2$SiO$_5$ crystals.

We here demonstrate a cavity-enhanced memory, where the crystal is placed in a impedance-matched cavity in order to achieve high absorption probability of a pulse incident on the cavity. The impedance-matched cavity approach was theoretically proposed in Refs \cite{Moiseev2010a,Afzelius2010a} and recently experimentally demonstrated in Refs \cite{Sabooni2013a,Sabooni2013} using a two-level AFC echo experiment, reaching up to 56\% efficiency. In those experiments the mirror coatings were applied directly on the REIC, a Pr$^{3+}$:Y$_2$SiO$_5$ crystal, making a very compact low-loss device. That approach, however, makes it extremely difficult to apply the optical control pulses required for optical-to-spin conversion, due to the sharp resonance of the cavity with respect to the frequency difference between the input/output and control pulses \cite{Sabooni2013a}. In principle a short ($\approx$1 cm) cavity would result in a cavity bandwidth large enough with respect to the relevant frequency separations (10 to 100 MHz for Eu$^{3+}$ and Pr$^{3+}$). But Sabooni \textit{et al.} \cite{Sabooni2013a} showed that the deep spectral holes which are the consequence of the necessary optical pumping in REIC crystals cause a steep dispersion, resulting in a strong reduction of the group velocity. The effect on the cavity is a reduction of the resonance linewidth, in some cases by several orders of magnitude \cite{Sabooni2013b}.

To solve this problem we have here chosen to build a longer cavity around the cryostat that cools the REIC, such that we can introduce control pulses in a crossed beam configuration, at the expense of the increased complexity due to the need of active cavity length stabilization of the cavity. In our experiment we observe an overall spin-wave AFC memory efficiency of 12\%, the highest reported so far. G\"{u}ndo\u{g}an et al. \cite{Gundogan2013} achieved 5.6\%, but in a material with much higher absorption coefficient. Our efficiency represents a 12-fold increase with respect to the efficiency we obtained in the experiment at the SPL \cite{Timoney2013}, which was our previous record for a Eu$^{3+}$:Y$_2$SiO$_5$ crystal. In comparison to the cavity-enhanced experiment in Ref. \cite{Sabooni2013}, where they obtained 56\% two-level AFC efficiency, we achieve a two-level AFC echo efficiency of 53\% for a comparable delay of 2 $\mu$s. In addition we also achieve the longest two-level AFC echo delays reached so far, up to 30 $\mu$s, which opens up the possibility to do multimode storage in the time domain.

The article is organised as follows. In section \ref{sec_AFC_theory} we give a short introduction to the principle of the impedance-matched AFC memory. In section \ref{sec_Experiment} we present the relevant properties of our REIC ($^{153}$Eu$^{3+}$:Y$_2$SiO$_5$), the optical pumping scheme, the laser system and memory cavity set-up. In section \ref{sec_Results} we present results of cavity-enhanced two-level AFC echoes and a full implementation of a AFC spin-wave memory. In section \ref{sec_Concl_Outlook} we give conclusions and an outlook.

\section{Cavity-enhanced atomic frequency comb memory}
\label{sec_AFC_theory}

The atomic frequency comb (AFC) memory \cite{Afzelius2009a} is a light-storage method developed for optical transitions having a static inhomogeneous broadening, which is the case for transitions in rare-earth-ion doped crystals \cite{Macfarlane1987,Macfarlane2002}. We thus consider an ensemble of atoms with an optical transition $|g\rangle-|e\rangle$ having a large inhomogeneous broadening $\Gamma_{in}$, but a narrow homogeneous linewidth $\gamma_h$ (i.e. $\Gamma_{in} \gg \gamma_h$). The many addressable spectral channels within the optical line allows a high-resolution spectral shaping of the $|g\rangle - |e\rangle$ transition by spectral hole burning, where an additional ground state $|aux\rangle$ is used as population storage reservoir. The optical transition is shaped into a periodic series of narrow peaks, of width $\gamma$, with periodicity $\Delta$, called the atomic frequency comb. The detailed experimental procedure for precise spectral shaping depends on the particular system, in section~\ref{sec_Experiment} we discuss the procedure for Europium-doped Y$_2$SiO$_5$ crystals. The light pulse to be stored should have a spectral bandwidth, $\gamma_p$, larger than the periodicity in the comb ($\gamma_p>\Delta$), but smaller than the total comb structure.

The interaction between an optical pulse and a ground-state population grating versus frequency generally results in a photon echo emission after a time $1/\Delta$, which is also the basis for accumulated or spectrally programmed photon echoes \cite{Hesselink1979,Carlson1984,Mitsunaga1991,Yano1992,Merkel1996,Tian2001a}. The echo emission arises from the evolution of the atomic coherence induced by the input pulse, which periodically rephases due to the periodicity in the atomic population grating. In typical echo experiments only a small fraction of the input pulse is re-emitted in the echo and the storage time is not variable since it is set by the predetermined grating periodicity $\Delta$. In Ref. \cite{Afzelius2009a} it is shown theoretically that a comb-shaped grating consisting of sharp and strongly absorbing peaks could generate a very efficient echo. In ref. \cite{Afzelius2009a} we considered Gaussian peaks, but Bonarota \textit{et al.} \cite{Bonarota2010} later realized that the optimal peak shape is squarish.

For a single-pass through the crystal, the memory efficiency can be expressed as \cite{Afzelius2009a}

\begin{equation}
\label{eta_nocavity}
\eta_{single-pass}=\tilde{d}^2 \exp(-\tilde{d}) \eta_{deph}
\end{equation}

\noindent where $\tilde{d}$ is the optical depth averaged over the photon bandwidth $\gamma_p$. For square peaks $\tilde{d}=d/F$, where $F$ is the comb finesse $F=\Delta/\gamma$. The factor $\eta_{deph}$ accounts for the dephasing due to the finite width of the peaks. For square peaks it can be written as $\eta_{deph}=\text{sinc}^2(\pi/F)$ \cite{Bonarota2010}. For a given optical depth $d$, there is an optimal comb finesse $F$. For square peaks it is given by \cite{Bonarota2010}

\begin{equation}
\label{F_opt}
F_{opt}=\frac{\pi}{\arctan (2\pi/d)}.
\end{equation}

\noindent  As an example one would require an optical depth of $d$=12, for which the optimal finesse is $F$=6.5, to achieve an efficiency of 50\%. Such optical depths are not possible to obtain with our present crystal, for a reasonable crystal length ($\approx$1 cm), where the peak optical depth $d$ is around 0.5-1.5, depending on the specific lambda system that is employed. We also emphasize that the efficiency of a single-pass memory, where the output echo is emitted in the same forward mode as the input pulse, the efficiency is bounded by 54\% due to the re-absorption of the echo (accounted for by the factor $\exp(-\tilde{d})$ in Eq. (\ref{eta_nocavity})).

To overcome the weak absorption of some REIC, it was proposed to use an optical cavity around the crystal and to operate it in a impedance-matched regime \cite{Moiseev2010a,Afzelius2010a}. The idea is to put the crystal in an asymmetric cavity, where the end mirror is totally reflecting and the front mirror has a reflectivity $R$. Total absorption of a pulse impinging on the cavity/crystal system is possible if the reflectivity is chosen as $R=\exp(-2\tilde{d})$, which is the impedance-match point. In this case the fraction of the internal cavity field leaking out of the front mirror has the same amplitude as the field reflected by the cavity, but a $\pi$ phase shift between the fields leads to total extinction of the reflection, hence the crystal absorbs all light. One can show that the memory efficiency is then only limited by the intrinsic dephasing of the memory \cite{Moiseev2010a,Afzelius2010a};

\begin{equation}
\label{eta_cavity}
\eta_{cavity}=\eta_{deph},
\end{equation}

\noindent provided that the cavity is loss-less, $\tilde{d} \ll 1$ and the input photon has a bandwidth much smaller than the cavity linewidth. Hence this configuration does not lead to re-absorption of light and the efficiency could approach the limit $\eta_{cavity}=1$ for low dephasing.

We will here analyse the effects of two imperfections that were not considered in the original proposals. First we take into account the finite value of $\tilde{d}$, whose effect can be derived using Eq. (11) in Ref. \cite{Afzelius2010a}, resulting in

\begin{equation}
\label{eta_cavity_dt}
\eta_{cavity}=\frac{\tilde{d}^2}{\text{sinh}^2(\tilde{d})} \eta_{deph}.
\end{equation}

\noindent This formula is valid under the assumption of a loss-less and impedance-matched cavity. As an example, for an average comb absorption of $\tilde{d}$=0.2, the theoretical efficiency is 0.99$\eta_{deph}$. The effect of a rather large $\tilde{d}$ is thus not very important. Secondly we take into account a finite loss $\epsilon$ (per cavity round trip). The effect of a loss can be derived using Eq. (14) in Ref. \cite{Afzelius2010a} and setting $R_2=1-\epsilon$, which results in

\begin{equation}
\label{eta_cavity_loss}
\eta_{cavity}=\frac{1}{\left(1+\frac{\epsilon}{4\tilde{d}}\right)^4} \eta_{deph},
\end{equation}

\noindent under the assumption that $\epsilon \ll \tilde{d} \ll 1$ and impedance matching. Losses is a much more serious limitation for weakly absorbing samples. For a $\epsilon/2$=0.01 single-pass loss and $\tilde{d}$=0.1, the efficiency would be bounded by $\eta_{cavity}$=0.82. This emphasizes the importance of low-loss optics and highly transparent host crystals.

We conclude this section by discussing the difference between a two-level AFC echo and a spin-wave AFC memory. The two-level AFC echo is a storage device with a pre-determined storage time, given by the prepared spectral grating. For a quantum memory it is essential that a variable storage time can be realized, so-called on-demand read out operation. A solution to this problem is based on coherent transfer of the excited state amplitude to a long-lived spin state before the appearance of the echo \cite{Afzelius2009a}, thus converting the optical coherence to a spin coherence. The memory can be read-out by transferring back the amplitude to the excited state, after a time determined by the user. We call this a spin-wave AFC memory, with emphasis on the true memory function. In addition this operation makes it possible to store the light for considerably longer durations. Spin-wave AFC memory experiments have only been demonstrated in Pr$^{3+}$:Y$_2$SiO$_5$ \cite{Afzelius2010,Gundogan2013} and Eu$^{3+}$:Y$_2$SiO$_5$ \cite{Timoney2012,Timoney2013} crystals. In principle the cavity enhancement of the input field has no influence on the spin-wave storage, but in practice it is clear that obtaining a cavity resonance for both the input and control field modes in a single spatial cavity mode would be challenging. Our solution to this problem is to send the control field off axis with respect to cavity mode.

\section{Experiment}
\label{sec_Experiment}

In this section we discuss the experimental set-up, which consists of a coherent laser source with associated light modulators, a REIC mounted in a low-vibration cooler and the cavity set-up around the cooler. We will also discuss some relevant crystal properties and the spectral hole burning schemes used to prepare the system.

\subsection{The crystal}
\label{sec_Cryst_prop}

\begin{figure}[ht]
    \centering
    \includegraphics[scale=0.7]{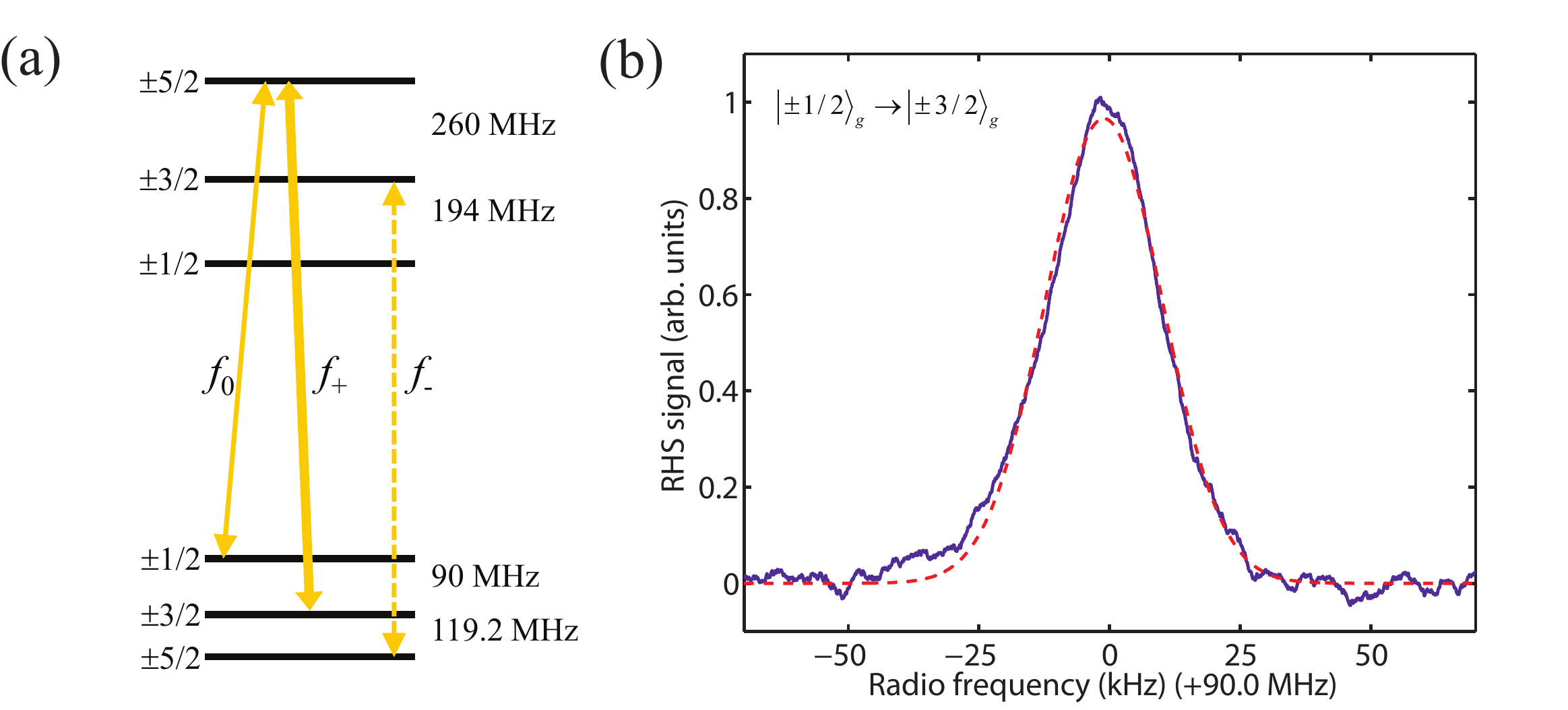}
    \caption{(a) Relevant energy-level diagram of $^{153}$Eu$^{3+}$ showing the optical-hyperfine transitions of the lambda-system for the AFC spin-wave memory, at frequencies $f_0$ and $f_+$, and the repump beam at frequency $f_-$. (b) Spin resonance spectrum of the $|\pm1/2\rangle_g-|\pm3/2\rangle_g$ transition at 90.0 MHz, which correspond to the inhomogeneous broadening profile of the spin ensemble used in the AFC spin-wave storage experiment. The solid blue line shows the spectrum obtained with the RHS technique (see text for details). The dashed red curve is a Gaussian fit to that line, where we obtained a FWHM linewidth of $\gamma_{spin}=26.5 \pm 0.1$ kHz.}
    \label{spin_resonance}
\end{figure}

The crystal is a isotopically pure $^{153}$Eu$^{3+}$:Y$_2$SiO$_5$ crystal, with 100 ppm $^{153}$Eu$^{3+}$ concentration. We use the yellow transition $^7$F$_0$ $\rightarrow$ $^5$D$_0$ at 580.04 nm of site 1, which has a extremely narrow homogeneous broadening \cite{Yano1991,Equall1994,Koenz2003} and a long spin coherence time \cite{Alexander2007}. At this doping concentration the optical transition has an inhomogeneous broadening of about 650 MHz and a peak absorption coefficient of $\alpha$=1.2 cm$^{-1}$ \cite{Lauritzen2012}, resulting in an optical depth $d=\alpha L$=1.2 for our $L$=1 cm crystal.

Several spectroscopic parameters relevant for quantum memories were recently investigated in $^{153}$Eu$^{3+}$:Y$_2$SiO$_5$ \cite{Lauritzen2012}, such as the hyperfine level order and the table of transition probabilities of all optical-hyperfine transitions for $^7$F$_0 \rightarrow ^5$D$_0$. In this work the input pulse excites the strong $|\pm1/2\rangle_g-|\pm5/2\rangle_e$ transition, at frequency $f_0$, and the control pulses excite the weaker $|\pm3/2\rangle_g-|\pm5/2\rangle_e$ transition, at frequency $f_+$, which are separated by $f_+-f_0$=90.0 MHz \cite{Yano1992a}. This is not the same lambda-system we used in our previous AFC spin-wave storage experiment with the $^{153}$Eu$^{3+}$ isotope \cite{Timoney2012}. The new lambda-system has a larger transition probability for the control pulse \cite{Lauritzen2012}, resulting in a higher control pulse Rabi frequency. In this work we also investigated the inhomogeneous spin broadening in more detail, which is relevant since it causes spin dephasing during the spin-wave storage time. An accurately measured spin line profile is relevant for estimating the different sources of inefficiencies in the final storage experiment (see section \ref{sec_Results_3LE}).

To obtain a spectrum of the $|\pm1/2\rangle_g-|\pm3/2\rangle_g$ spin transition at 90.0 MHz we used the Raman-heterodyne scattering (RHS) technique \cite{Mlynek1983,Wong1983}. We first initialize all ions in the $|\pm1/2\rangle_g$ state (see section \ref{sec_Prep_method}). Then the spin transition is excited with a RF field while the optical  $|\pm3/2\rangle_g-|\pm5/2\rangle_e$ is excited by an optical field. In this way the coherence on the spin transition induced by the RF excitation is transferred into an optical coherence on the $|\pm1/2\rangle_g-|\pm5/2\rangle_e$ transition, which in turn results in coherent emission on this transition in the same spatial mode as the optical excitation beam. This emission is then detected as a beat signal on the optical transmission signal, which can be detected using photo detectors and de-modulation techniques. The recorded spin absorption profile is shown in Figure \ref{spin_resonance}. The profile fits well to a Gaussian function with a FWHM linewidth $\gamma_{spin}=26.5 \pm 0.1$ kHz. The spin broadening on this transition is significantly lower than on the $|\pm3/2\rangle_g-|\pm5/2\rangle_g$ at 119.2 MHz, which we previously measured to be $\gamma_{spin}=69 \pm 3$ kHz by fitting the decay of the efficiency as a function of spin-wave storage time \cite{Timoney2012}. The narrower spin linewidth provides another advantage of the new lambda system. Note also that the shape and width of the spin resonance can be measured with an increased accuracy using the RHS technique.

\subsection{The optical pumping pulse sequence}
\label{sec_Prep_method}

REIC crystals often require optical pumping sequences in order to prepare an ensemble of ions which can be excited on specific transitions, in our case the lambda system that we will use for spin-wave memory operations. The reason for this is that the inhomogeneous broadening normally is larger than the hyperfine spacings in the ground and excited states. For Eu$^{3+}$ and Pr$^{3+}$ ions, with three hyperfine levels in the ground and excited states, it means that without special care a given optical frequency would be in resonance with 9 different optical-hyperfine transitions, which defines 9 classes of ions.

The first preparation step, called class cleaning, solves this problem by optically pumping away ions from all but one class. As explained in Ref. \cite{Lauritzen2012} we can easily achieve this by simultaneously optically pump with frequencies $f_0$, $f_+$ and a third frequency $f_-$, in our case the transition  $|\pm5/2\rangle_g-|\pm3/2\rangle_e$ shifted by $f_{-}-f_0$=-51 MHz. The second step is to spin polarize all ions of that selected class into the ground state $|\pm1/2\rangle_g$ by applying pulses at $f_+$ and $f_-$. During first and second steps we also sweep all frequencies over 15 MHz, such that we have a well-defined ensemble spin polarized over this frequency range. The peak optical depth on the $f_0$ transition after the spin polarization is $d$=0.8. Steps 1 and 2 take 300 ms. 

The third step is to create the comb on the $f_0$ transition, which we do by frequency-selectively pumping away ions on that transition. Each comb creation pulse is 500 $\mu$s long. After each comb creation pulse follows a 500 $\mu$s long pump pulse on the transition $f_+$ to keep state $|\pm3/2\rangle_g$ empty of population. Ions not used in the comb are thus pumped into state $|\pm5/2\rangle_g$. This sequence is repeated 300 times. Note that the excited population decays with the time constant $T_1$=2 ms. Several methods for preparing combs have been investigated using both amplitude and frequency modulation \cite{Riedmatten2008,Sabooni2010,Clausen2011,Saglamyurek2011,Bonarota2010}. As discussed in section \ref{sec_AFC_theory} the optimal peak shape is squarish, which is challenging to obtain. We here use a sequence of complex hyperbolic secant pulses (sech pulse) \cite{Rippe2005} to optically pump away ions in trenches with as sharp edges as possible. If the pulse parameters are properly chosen these pulses can have very squarish spectral transfer functions. The adiabacity of the pulses also make them less sensitive to experimental imperfections. To define a comb the 500 $\mu$s pump pulse is a sum of $N_{peak}$ sech pulses, each centred on a different frequency $f_0+m \Delta$, where $m$ is an integer. The number of frequency bands corresponds to the number of peaks created $N_{peak}$. The total bandwidth of our comb was about 5 MHz, such that $N_{peaks}$=5MHz/$\Delta$. This complicated pump pulse can be created by directly feeding a corresponding radio-frequency signal into an acousto-optic modulator (AOM). The details of this pump sequence, its technical implementation and comparisons to other pumping techniques will be presented in a separate paper \cite{Jobez2014}.

\subsection{The laser system and cavity set-up}
\label{sec_Laser_Cavity_setup}

\begin{figure}[h]
    \centering
    \includegraphics[scale=0.6]{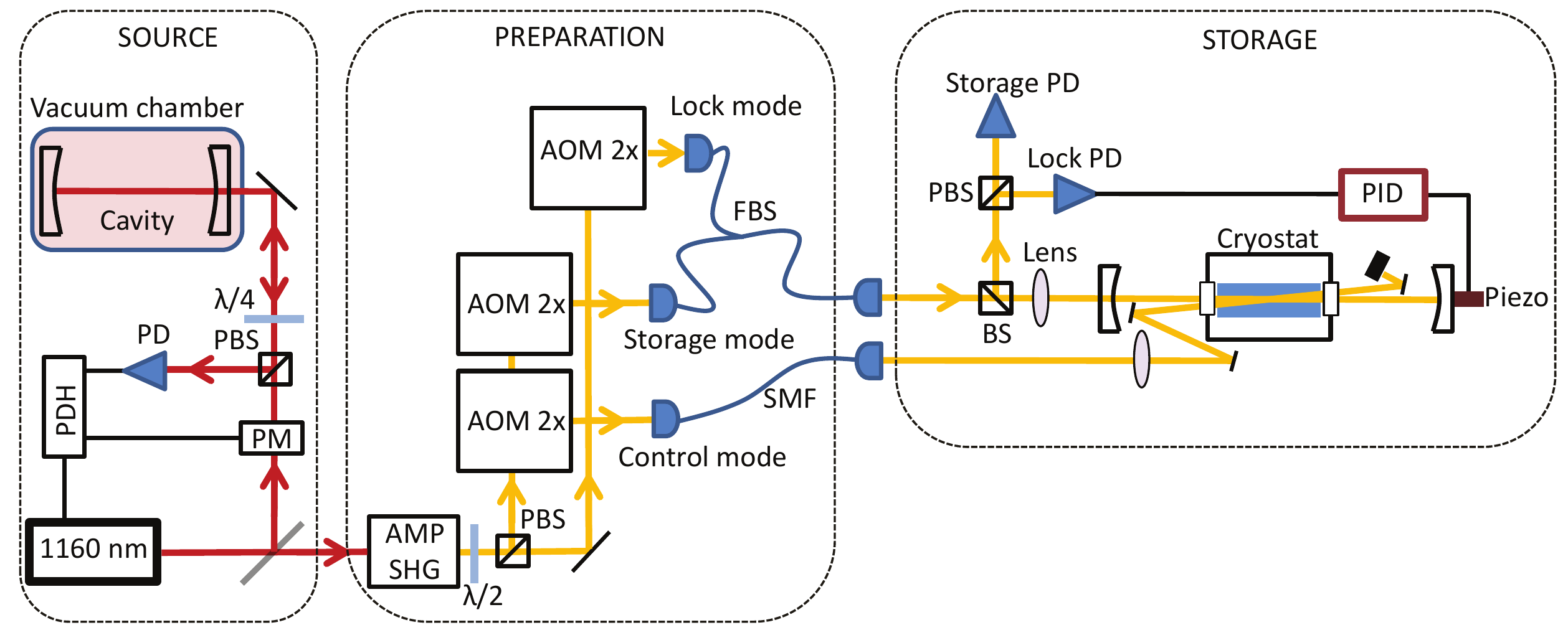}
    \caption{The experiment consists of a laser source (source), light modulators for optical pumping (preparation) and the cavity-enhanced memory (storage). The 1160 nm diode laser source (Toptica DL Pro) is split by a 10/90 mirror, where the weaker beam is sent to a laser stabilization set-up and the stronger beam is amplified using a fiber Raman amplifier (AMP) and frequency doubled (SHG) to 580 nm (using a MPBC VRFA device). The laser is stabilized on a narrow linewidth cavity ($\approx$ 100 kHz) placed in a vacuum chamber using the Pound-Drever-Hall (PDH) technique \cite{Black2000}. The 580 nm light is sent to three AOMs which are operated in a double-pass configuration (polarization-based) to provide individual amplitude and frequency control of the storage, control and lock modes. All modes are coupled into single-mode fibers (SMFs), the lock and storage modes are combined into one mode using a fiber beam splitter (FBS) with orthogonal polarizations (using fibered polarization controllers not shown). The storage/lock modes are coupled into the memory cavity using a lens and the reflected light is detected after the BS. The storage and lock modes are split with a PBS and detected on separate photodiodes (PDs). The cavity length is stabilized with the lock mode using a PID and a piezo. The control mode is focused into the cavity with a separate lens and crosses the storage mode inside the cavity. The control mode is the most intense, with 500 mW peak power before the cryostat. The crystal is mounted in a low-vibration cooler (Cryostation, Montana Instruments) cooling the crystal to 3 K. ($\lambda/4$=quarter-wave plate,$\lambda/2$=half-wave plate,PM=phase modulator)}
    \label{exp_setup}
\end{figure}

The experimental set-up is shown in Figure \ref{exp_setup}. The laser source is a amplified and frequency-doubled 1160 nm diode laser producing 580 nm. The 1160 nm seed laser is frequency stabilized by locking it to an optical cavity. The short-term ($\le$1ms) frequency linewidth of the system was evaluated by doing optical free-induction experiments in a Eu$^{3+}$:Y$_2$SiO$_5$ crystal, resulting in an average linewidth of a few kHz. We have however not yet been able to properly evaluate the long-term linewidth. The 580 nm laser beam is split into two spatial modes, the storage and control mode, both which are controlled by double-pass AOMs. Both modes are sent through optical fibers after the AOMs for spatial mode cleaning. 

The memory set-up consists of the crystal mounted in a low-vibration 3-K cooler, and the optical cavity around the cooler. The cavity is made up of two mirrors with reflectivity $R_1$=0.73 and $R_2$=0.995, having radius of curvature $R_C$=0.15 m. The cavity is operated close to the concentric limit with cavity length $L=2R_C$ where the cavity mode is tight. Estimating the cavity mode waist is difficult due to its non-linear dependence on $L$ close to the $L=2R_C$ point. By measuring the transverse mode spacing we estimate it in the range 60-90$\mu$m. We can then also use a tight control mode to maximize the Rabi frequency for the control pulses, which we set to 145$\pm$10 $\mu$m. The free spectral range (FSR) is close to 500 MHz and the cavity finesse is 20. The input mode is sent through a 50/50 beam-splitter (BS) and focussed in the middle of the cavity. The light reflected by the cavity is detected in the mode reflected by the BS. This results in an additional 25\% device loss, which could be removed by using a Faraday rotator and a polarizing BS (PBS) to separate the input and output modes. Note that the reflected light is sent directly onto the photo diode, without passing through an optical fiber. The control mode is focussed through the crystal with an angle, without passing through the cavity mirrors. Both modes are polarized along the crystal D$_1$ axis to maximize the transition dipole moment. All optical pumping pulses and the control pulses where applied in the control mode, only the input pulse was sent through the storage mode.

The crystal surfaces and the two cooler windows where anti-reflected (AR) coated to reduce losses. The total losses for a single pass through the cooler where roughly $\epsilon/2 \approx$ 0.015, meaning 0.2\% losses per surface (6 in total). The impact of the losses on the results will be discussed in section \ref{sec_Results_2LE}.

As mentioned in the introduction, the cavity resonance narrows due to the strong dispersion in the 15 MHz wide transmission window around the $f_0$ frequency, as first observed in Ref. \cite{Sabooni2013a}. We measured that the cavity linewidth in the transmission window was about 3 MHz, while outside the absorption line it was 25 MHz. The narrowing is in accordance with the theoretical prediction \cite{Sabooni2013a}, which is 2 MHz, given the parameter uncertainties.

To maintain the memory cavity in resonance with the $f_0$ frequency, we designed an active side-of-fringe locking scheme. A third laser beam (lock mode) is derived from the laser, which is shifted one FSR from $f_0$. The lock beam is orthogonally polarized with respect to the storage mode, such that it can be separated from the storage mode using a PBS. The locking scheme is intermittent, meaning a lock signal is generated only during the AFC preparation sequence. The feedback is done using a piezo on which the $R_2$ mirror is mounted. 

\section{Results and Discussion}
\label{sec_Results}

We here present the experimental results obtained in this work. We start by showing the results of cavity-enhancement of the two-level AFC echoes. We then show the results of a full implementation of the AFC quantum memory, with a cavity-enhanced memory efficiency of 12\%.

\subsection{Cavity-enhanced two-level AFC echo}
\label{sec_Results_2LE}

\begin{figure}[ht]
    \centering
    \includegraphics[width=1\textwidth]{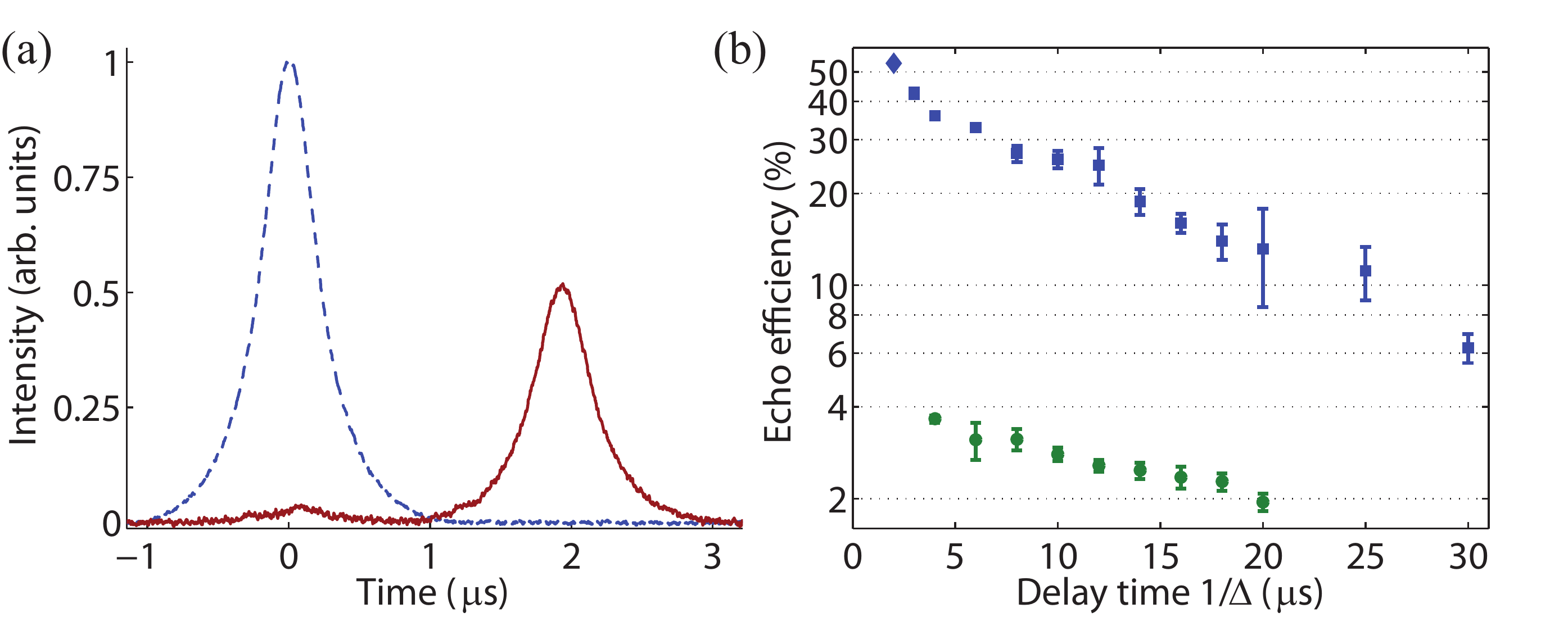}
    \caption{In (a) we show the highest two-level AFC echo efficiency obtained in this work, which was 53$\pm$1\%. The dashed (blue) curve is the input pulse, the solid (red) curve shows the echo at $1/\Delta$=2$\mu$s. In (b) we show the two-level AFC echo efficiency as a function of the inverse comb spacing $1/\Delta$. The circles (green) show the efficiency obtained without cavity for a single pass through the crystal. The squares (blue) show the efficiency for the cavity-enhanced memory. The diamond shows the efficiency for the case shown in panel (a), where the input pulse duration was significantly shorter (450 ns). For all other measurements the input pulse had an approximately Gaussian shape with a FWHM of 1.5 $\mu$s. The efficiencies where calculated in terms of the area of the pulses. The error bars are the standard deviation of about 5 measurements.}
    \label{2LE_comparison}
\end{figure}

The highest two-level efficiency we obtained was $\eta_{2L}$=53\%, for a delay of $1/\Delta$=2$\mu$s, see Fig. \ref{2LE_comparison}a.  This is close to the maximum efficiency obtained by Sabooni \textit{et al.} \cite{Sabooni2013} (56\%) for a delay of $1/\Delta$=1.1$\mu$s. Several factors limit the experimentally feasible efficiency in our case. We first consider a theoretical limit set by the optical peak depth $d$=0.8 and the mirror reflectivity $R_1$=0.73. The mirror reflectivity implies that the impedance match is obtained for $\tilde{d}$=0.16, which means that the optimal comb finesse for our system would be $F=d/\tilde{d}=$5. For square peaks that means a maximum theoretical efficiency of $\eta_{cavity}=\text{sinc}^2(\pi/F)=$0.88. There are also two technical limitations due to the losses in the cavity and mode matching between the storage and cavity modes. The cavity losses of $\epsilon/2=$0.015 sets a maximum efficiency of 0.83 (see Eq. (\ref{eta_cavity_loss})), for $\tilde{d}$=0.16. In combination with the optimal comb finesse the efficiency would be bounded by 0.88$\cdot$0.83=0.73. The spatial mode matching error is difficult to estimate, but we note that experimentally it was challenging to obtain a sufficiently good spatial mode matching. The difference between the estimated 73\% efficient memory and the observed 53\% record efficiency is probably due to spatial mode matching and, to a lesser degree, a non-perfect comb preparation in terms of shape, finesse and peak optical depth $d$.

The efficiency decreases, however, for longer delays. In Fig. \ref{2LE_comparison}b we present two series of measurements showing $\eta_{2L}$ as a function of the inverse comb spacing $1/\Delta$, without and with cavity enhancement. To obtain the data without cavity enhancement, we simply inserted a mirror before the 100\% reflecting cavity mirror, which directed the light onto a detector (single pass through crystal). The input pulse trace is obtained by creating a transparent window around the input pulse frequency. For the case of cavity enhancement, where we detect the light reflected off the cavity, we measured the input pulse trace by simply inserting a block after the partly reflecting cavity mirror and by normalizing with the known mirror reflectivity. The data shown in Fig. \ref{2LE_comparison}b results in a cavity-enhanced efficiency of 10 for some of the delays, with an average gain of 8.4. We also note that the single-pass efficiency $\eta_{single-pass}$=3.7$\pm$0.1\% obtained at $1/\Delta$=4$\mu$s is close to the theoretical limit that is 4.4\% for $d=0.8$ (assuming square peaks).

In addition to the strong cavity enhancement, we also emphasize on the long $1/\Delta$ delays obtained in this work. Previous AFC echo experiments have reached delays up to about 5 $\mu$s at most \cite{Timoney2013,Gundogan2013}, with a rapid decay for delays beyond this value. Here we observe two-level efficiencies above 10\% up to  $1/\Delta$=25$\mu$s, corresponding to a teeth separation of $\Delta$=40kHz. A long $1/\Delta$ delay is crucial since it allows to store more temporal modes \cite{Afzelius2009a}, which in turn will be necessary for some applications such as quantum repeaters \cite{Simon2007}. Several difficulties where encountered when attempting to create high-resolution combs with high peak absorption. Some were technical, such as laser frequency linewidth and frequency noise due to AOM drivers, while others were more fundamental, such as optimizing the frequency-selective optical pumping sequence used to create the comb structure with optimal square teeth, as briefly discussed in section \ref{sec_Prep_method}. The reason for the decay remains, however, unknown. It could be due to several factors, for instance laser frequency noise, imperfections in the implementation of the preparation sequence or the homogeneous linewidth of the material (including spectral diffusion during the long comb preparation period). This will be investigated in future studies.

\subsection{Cavity-enhanced spin-wave AFC memory}
\label{sec_Results_3LE}

\begin{figure}[h]
    \centering
    \includegraphics[scale=0.75]{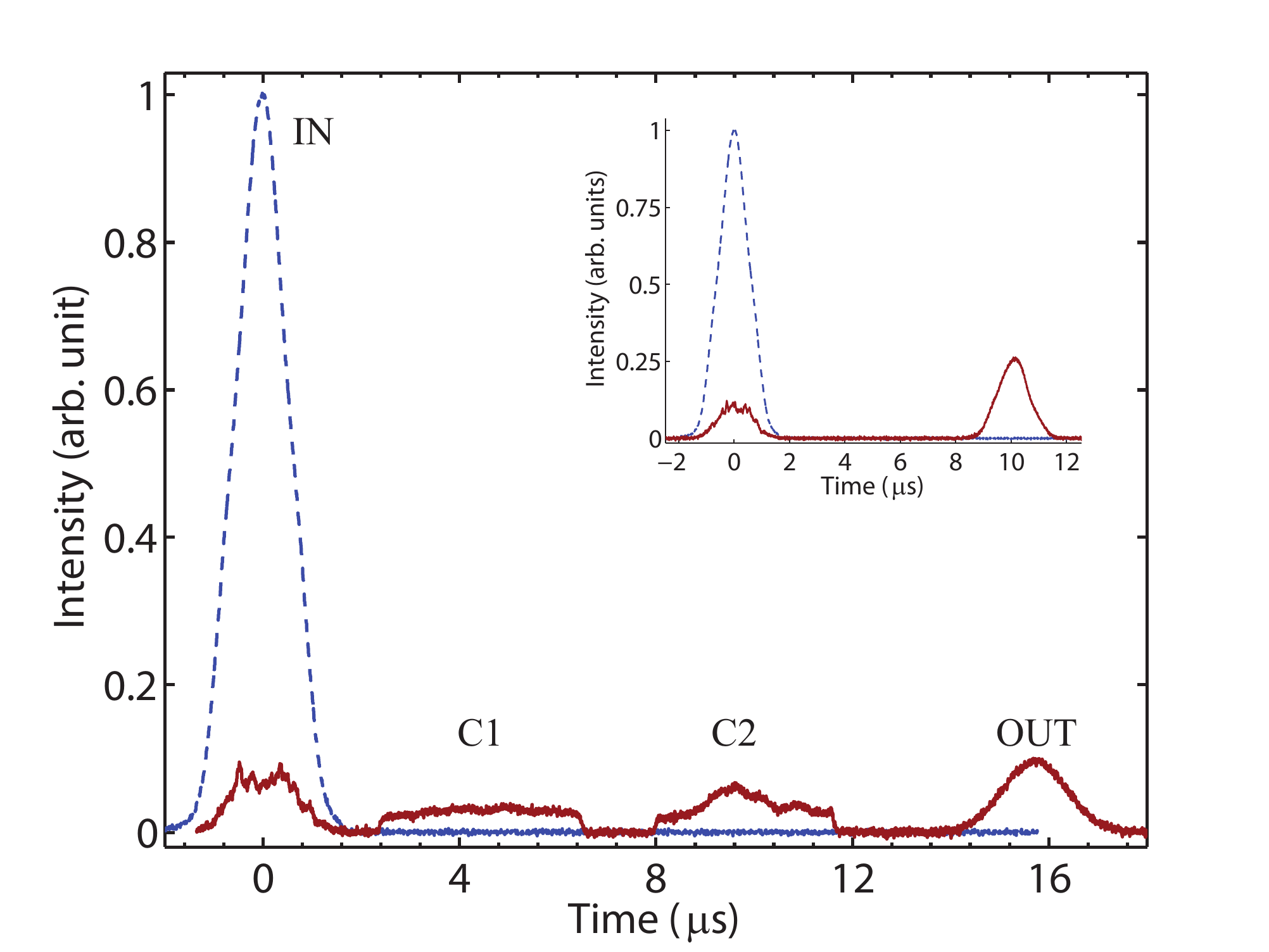}
    \caption{Experimental example of a cavity-enhanced spin-wave AFC memory. The dashed (blue) line shows the input reference signal (IN). The solid (red) line shows the detected signal when applying the control pulses, with the memory output signal (OUT) being visible at about 16 $\mu$s. One can also see scattered light from the strong control pulses (C1 and C2). The output signal area is 12\% of that of the input signal area. We note that the output signal is approximately 1.2 times longer than the input, probably owing to a spectral truncation due to the finite bandwidth of the control pulses. The shape of the first control pulse is representative for the control pulses (see discussion in text on the pulse shape), while the second pulse interferes with the residual two-level AFC echo at 10 $\mu$s. Inset: The 28\% efficient two-level AFC echo at 10 $\mu$s obtained without applying the control pulses.}
    \label{spin_wave_memory_example}
\end{figure}

We now turn to the main result of this work, a cavity-enhanced spin-wave AFC memory. In Figure \ref{spin_wave_memory_example} we show the highest spin-wave AFC memory efficiency obtained in this work, $\eta$=12\%. For this measurement we programmed an AFC delay of $1/\Delta$=10$\mu$s and the spin-wave storage time was set to $T_{sw}$=5.3 $\mu$s, defined as the distance between the centre of the two control pulses. The two-level AFC echo efficiency, without applying the control pulses, was measured to be 28\% (see inset in Fig. \ref{spin_wave_memory_example}), which is consistent with the measurements shown in Fig. \ref{2LE_comparison}b. 

The main difficulty in obtaining this result is the finite time that is required to apply the optical control pulses that transfer the optical excitation to the spin state and back. We obtained a maximum Rabi frequency of 250 kHz for the control mode, meaning that a simple $\pi$ pulse would have a duration of 2 $\mu$s. A $\pi$ pulse is not optimal, however, since only the carrier frequency in the spectrum achieves unit transfer efficiency. To obtain a more uniform spectral transfer function one can use adiabatic pulses \cite{Minar2010}, for instance the sech pulses that we also use for the comb creation (see section \ref{sec_Prep_method}). The sech pulse amplitude envelope has the form $\Omega(t)=\Omega_{max} \text{sech}(1.76t/T)$, where $\Omega_{max}$ is the peak Rabi frequency and $T$ is the FWHM pulse duration in intensity. The pulse is also frequency chirped as $\nu(t)=\delta\nu/2 \tanh(1.76t/T)$, where $\delta\nu$ is the total frequency chirp range. If the adiabatic conditions are fulfilled, see Ref. \cite{Rippe2005}, a sech pulse can reach unit transfer efficiency over the entire bandwidth $\delta\nu$. In our work we only used partially adiabatic pulses, indeed by varying all pulse parameters to optimize the spin-wave memory efficiency we found that truncated pulses gave the best overall efficiency. This is probably due to the larger pulse area of the truncated pulses. The sech control pulses had parameters $T$=5$\mu$s and $\delta\nu$=1.2 MHz, with an estimated $\Omega_{max}$=250 kHz. The pulses where truncated to a duration of 4$\mu$s, which gives the squarish pulses with a smooth top that can be seen in Figure \ref{spin_wave_memory_example}. 

To gain further insight into the efficiency of the memory, we performed a detailed analysis of the limiting factors of the obtained efficiency. The spin-wave memory efficiency is given by $\eta=\eta_{2L} \eta_{T}^2 \eta_{sw}$, where $\eta_{T}$ is the transfer efficiency of one control pulse and $\eta_{sw}$ is the dephasing due to the inhomogeneous spin linewidth. The exact form of the spin dephasing depends on the shape of the inhomogeneous spin linewidth. In previous work a Gaussian spin lineshape was assumed \cite{Gundogan2013,Afzelius2010,Timoney2012,Timoney2013}, but without an independent experimental verification of that assumption. To obtain a more accurate measure of the lineshape, which allows a more accurate evaluation of the control pulse efficiency $\eta_{T}$, we performed an independent spectroscopic measurement of the spin transition. As explained in section \ref{sec_Cryst_prop} we measured a Gaussian spin broadening of 26.5 kHz using the RHS technique. For a Gaussian shape the spin dephasing is given by $\eta_{sw}(T_{sw})=\exp(-\gamma_{spin}^2 T_{sw}^2 \pi^2/(2 \ln(2)))$ \cite{Timoney2012}. For the parameters above we then get $\eta_{sw}(5.3 \mu s)=0.87$, resulting in a spin-wave memory efficiency of 14\% when extrapolated to $T_{sw}$=0 $\mu$s. From this analysis it further follows that the control pulse efficiency is $\eta_{T}$=0.70. This analysis is based on a simplified single-mode model that neglects the spatial overlap between the modes. We also performed numerical calculations of the control pulse efficiency by numerically solving the Bloch equations \cite{Burr2004}, which is also based on a spatial single-mode model, resulting in an efficiency of 0.91. The difference is probably due to the neglected transverse profile of the spatial modes and non-perfect overlap in the crossed beam configuration. 

There is likely room for improvement in the optimization of the spatial modes and overlap. For instance, a higher Rabi frequency in the control mode could be achieved by focusing harder, but that would require a redesign of the cavity, since its parameters impose the mode size of the storage mode. We also emphasize that the inefficiency due to the spin dephasing could in principle be eliminated using two radio-frequency $\pi$-pulses, that is a spin echo sequence. This would also lead to longer spin-wave storage times, since the spin coherence time is longer than 10 ms \cite{Alexander2007} in this crystal. 

\section{Conclusions and Outlook}
\label{sec_Concl_Outlook}
We have demonstrated cavity-enhanced storage in an optical AFC spin-wave memory. For the two-level AFC echo we reached storage efficiencies up to 53\% for short delays (2 $\mu$s), comparable to previous results \cite{Sabooni2013}. In addition we demonstrate efficient two-level AFC echoes for unprecedented long delays, with efficiencies above 10\% up to 25 $\mu$s delay. The main new result reported here is cavity-enhancement of an AFC spin-wave memory, where we reached 12\% efficiency. This result is in particular important for intrinsically low-absorbing materials, such as the Europium-doped crystal used in this work. We have also identified several possible improvements that could further boost the efficiency. For instance, a more careful design of the spatial control and cavity modes could lead to a higher transfer efficiency by the optical control beams, or the use of optics with even lower losses could further reduce parasitic losses. Our result should make it possible to significantly improve on current state-of-the-art results when storing optical pulses at the single-photon-level, where we were hampered by a low storage efficiency \cite{Timoney2013}.

\section{Acknowledgements}

We thank Nicolas Sangouard, F\'{e}lix Bussi\`{e}res, Hugues de Riedmatten and Sergey Moiseev for useful discussions and Claudio Barreiro for technical support. We gratefully acknowledge R. Cone and R. M. MacFarlane for lending us the isoptically pure $^{153}$Eu:Y$_2$SiO$_5$ crystal. This work was financially supported by the Swiss National Centres of Competence in Research (NCCR) project Quantum Science Technology (QSIT) and by the European projects SIQS (FET Proactive Integrated Project) and CIPRIS (People Programme (Marie Curie Actions) of the European Union’s Seventh Framework Programme FP7/2007-2013/ under REA Grant No. 287252).

\section{References}
\providecommand{\newblock}{}

\bibliographystyle{iopart-num}

\end{document}